\documentclass[prb,aps,twocolumn,showpacs,showkeys]{revtex4}
\usepackage[english]{babel}
\usepackage{epsfig}
\usepackage{graphicx}
\usepackage{amsmath,amssymb,amsfonts}
\usepackage{bm}

\begin{document}

\title{Jahn-Teller distortions and phase separation in doped
manganites}

\author{A.\,O.~Sboychakov, K.\,I.~Kugel, and A.\,L.~Rakhmanov}
\affiliation{Institute for Theoretical and Applied
Electrodynamics, Russian Academy of Sciences, Izhorskaya Str.
13/19, Moscow, 125412 Russia}

\begin{abstract}
A ''minimal model'' of the Kondo-lattice type is used to describe
a competition between the localization and metallicity in doped
manganites and related magnetic oxides with Jahn-Teller ions. It
is shown that the number of itinerant charge carriers can be
significantly lower than that implied by the doping level $x$. A
strong tendency to the phase separation is demonstrated for a wide
range of intermediate doping concentrations vanishing at low and
high doping. The phase diagram of the model in the $x-T$ plane is
constructed. At low temperatures, the system is in a state with a
long-range magnetic order: antiferromagnetic (AF), ferromagnetic
(FM), or AF--FM phase separated (PS) state. At high temperatures,
there can exist two types of the paramagnetic (PM) state with zero
and nonzero density of the itinerant electrons. In the
intermediate temperature range, the phase diagram includes
different kinds of the PS states: AF--FM, FM--PM, and PM with
different content of itinerant electrons. The applied magnetic
field changes the phase diagram favoring the FM ordering. It is
shown that the variation of temperature or magnetic field can
induce the metal-insulator transition in a certain range of doping
levels.
\end{abstract}



\date{\today}

\maketitle

\section{Introduction}

The effect of electron correlations on the properties of different
materials is currently among the most burning problems of the
condensed matter physics. As a rule, strong electron correlations
are accompanied by the formation of nanoscale inhomogeneous
states~\cite{DagSci}. Such inhomogeneities have been already
studied for several decades. In particular, they were widely
discussed for high-$T_c$ superconductors~\cite{NagSup} and heavy
fermion materials~\cite{fermion}. In the recent years, they
attract a special attention owing to the discovery of the colossal
magnetoresistance effect in manganites (the nature of which is
believed to be closely related to inhomogeneous
structures~\cite{dagbook}). The inhomogeneities manifest
themselves in other magnetic materials such as
cobaltites~\cite{cobalt}, nickelates~\cite{nickel} and also in
low-dimensional magnets~\cite{low-d}. All these systems are
characterized by a strong interplay of spin, charge, and orbital
degrees of freedom leading to rather rich phase diagrams.

One of the first most spectacular examples of such kind of
inhomogeneities is the formation of ferromagnetic (FM) droplets
(magnetic polarons or ferrons) in antiferromagnetic (AF)
semiconductors at low doping levels as well as FM spin polarons in
the paramagnetic state~\cite{Nag67,Kasuya}. These examples
correspond to the case of so called electron phase separation
caused by self-trapping of charge carriers, which change their
local environment. In addition to such a small-scale phase
separation, in manganites, as well as in other compounds
exhibiting first order transitions (e.g., between FM and AF
phases), there also arises the phase separation of another type
related to rather wide region where different phases coexist. An
example of such large-scale phase separation is the formation of
rather large FM droplets with the size of the order of (100-1000)
$\AA$ inside the AF matrix~\cite{Bala,Ueh}. At higher doping
levels close to the half-filling, there appears one more threshold
for the phase separation in the system corresponding again to the
formation of ferromagnetic droplets, but now in a charge-ordered
insulating matrix~\cite{KKK}. The interaction of spin, charge, and
orbital degrees of freedom can also lead to the formation of
stripe structures instead of droplets at high content of the
alkaline-earth element~\cite{stripes}. In manganites, owing to the
strong electron-lattice interactions, such structures are related
to the lattice distortions and can be observed by the electron
diffraction and the low-angle neutron scattering~\cite{KuKh}.

Both analytical and numerical studies in various models related to
the strongly correlated electrons exhibit a pronounced tendency
toward phase separation in a wide range of temperatures and
electron or hole concentrations. Among these, we can mention
$s-d$~\cite{Kak}, $t-J$~\cite{t-J}, Hubbard~\cite{Viss}, and
Falicov-Kimball~\cite{PSfal} models.

The theoretical models usually imply that the number of charge
carriers introduced by doping is equal to the number of itinerant
electrons which take part in the formation of nanoscale
inhomogeneities. However, the comparison of experimental data with
the theoretical results suggests that such an approach is
insufficient~\cite{zhao} and the number of self-trapped carriers
can significantly differ from the doping
level~\cite{prbMi,zhetf04}.

Here, we analyze the model proposed in our paper
Ref.~\onlinecite{prlKRS}, which relates the doping level and the
number of charge carriers. We take into account the Jahn-Teller
(JT) nature of magnetic ions, which could give rise to the
localization of charge carriers at the lattice distortions. We
introduce this localization effect to the Kondo-lattice model in
the double exchange limit. Beginning from seminal paper,
Ref.~\onlinecite{Millis}, the role of JT distortions was widely
discussed in literature~\cite{dagbook,hor,bus}. In particular,
such distortions were taken into consideration in
Refs.~\onlinecite{Rama,Rama1} in the analysis of the phase diagram
of doped manganites. However, none of these papers dealt with the
phase separation phenomena. In our paper, the main emphasis is
made on the studying the phase diagram in the doping--temperature
($x-T$) plane and its evolution under effect of the applied
magnetic field.

\section{The model}

First, let us note that the doped manganites,
(Re$_{1-x}^{3+}$Mn$_{1-x}^{3+}$)(A$_{x}^{2+}$Mn$_{x}^{4+}$)O$_3^{2-}$,
are the compounds with mixed valence. Here Re is a trivalent
rare-earth element and A is a bivalent alkaline-earth element. The
ions Mn$^{3+}$ and Mn$^{4+}$ have $3d^{4}$ and $3d^{3}$ electron
configurations, respectively. In such compounds, manganese ions
are located in the centers of O$_6$ octahedra. In a regular
octahedron, a five-fold degenerate $3d$ level is split into
triple- and double-degenerate levels, $t_{2g}$ ($d_{xy}$,
$d_{yz}$, $d_{zx}$) and $e_{g}$ ($d_{x^2-y^2}$, $d_{3z^2-r^2}$),
respectively. The $t_{2g}$ level lies lower than $e_{g}$ level.
Manganese ions are characterized by the strong Hund's rule
coupling giving rise to the parallel alignment of intra-atomic
electron spins. So, the spins of $t_{2g}$ electrons form local
spin $S=3/2$. In a regular MnO$_6$ octahedron, Mn$^{3+}$ ion has
one electron at double-degenerate $e_{g}$ level~\cite{Good}.
According to  the Jahn-Teller theorem, the latter configuration is
unstable and the degeneracy is lifted by a distortion of the
octahedron. The Mn$^{4+}$ is not a Jahn-Teller ion and
Mn$^{4+}$O$_6$ octahedron remains undistorted. The distortion of
Mn$^{3+}$O$_6$ octahedron leads to the energy lowering by
$\epsilon_{JT}$. In doped manganites, the $e_{g}$ electron can hop
from Mn$^{3+}$ to Mn$^{4+}$ ion producing a gain in the kinetic
energy due to electron delocalization. Therefore, an electron can
lower its energy either due to the Jahn-Teller-induced
localization at distorted octahedra or by the delocalization
related to the inter-atomic hopping. The strong Hund's rule
coupling favors the hopping of an electron when its spin is
parallel to the spin of core ($t_{2g}$) electrons. This is the
origin of the well-known double-exchange mechanism of
ferromagnetic interaction between the localized
spins~\cite{Zener}.

Therefore, it is natural to assume that the $e_g$ electron can be
either localized due to Jahn-Teller distortions with the energy
gain $\epsilon_{JT}$ ($l$ electron) or to decrease its energy due
to band broadening ($b$ electron)~\cite{Pai,Rama,Rama1}. So, there
exists a competition between the localization and delocalization.
Such a system with localized and band electrons can be analyzed
using the Hubbard Hamiltonian taking into account the
electron-lattice interaction, the Hund's rule coupling, and the
exchange interaction between core electrons~\cite{prlKRS}
\begin{eqnarray}\label{H}
\hat{H}&=&H_{\text{el}}+H_{\text{AF}}+H_{\text{JT}}+H_{\text{el-el}}\,,\\
H_{\text{el}}&=&-\sum_{\langle{\bf n}{\bf
m}\rangle}\sum_{ab\sigma}\left(t^{ab}_{{\bf n}{\bf
m}}a^{\dag}_{{\bf n}a\sigma}a_{{\bf
m}b\sigma}+h.c.\right)\nonumber\\&&-\frac{J_H}{2}\!\sum_{{\bf
n}}\!\sum_{a\sigma\sigma'}a^{\dag}_{{\bf
n}a\sigma}\left({\bm{\sigma}}{\bf S}_{{\bf
n}}\right)_{\sigma\sigma'}a_{{\bf n}a\sigma'}\,,\nonumber\\
H_{\text{AF}}&=&J'\!\!\!\sum_{\langle{\bf n}{\bf m}\rangle}{\bf
S}_{{\bf n}}{\bf S}_{{\bf m}},\nonumber\\
H_{\text{JT}}&=&-g\sum_{{\bf n}}\sum_{ab\sigma}a^{\dag}_{{\bf
n}a\sigma}\left[Q_{2{\bf n}}(\sigma^{x})_{ab}+Q_{3{\bf
n}}(\sigma^{z})_{ab}\right]a_{{\bf n}b\sigma}
\nonumber\\&&+\frac{K}{2}\sum_{{\bf
n}}\left(Q_{2{\bf n}}^{\,2}+Q_{3{\bf n}}^{\,2}\right)\,,\nonumber\\
H_{\text{el-el}}&=&\frac{U_1}{2}\sum_{{\bf n}a\sigma}n_{{\bf
n}a\sigma}n_{{\bf n}a\bar{\sigma}}+\frac{U_2}{2}\sum_{{\bf
n}a\sigma\sigma'}n_{{\bf n}a\sigma}n_{{\bf
n}\bar{a}\sigma'}\,.\nonumber
\end{eqnarray}

In this Hamiltonian, $a^{\dag}_{{\bf n}a\sigma}$, $a_{{\bf
n}a\sigma}$ are creation and annihilation operators for $e_{g}$
electrons at site ${\bf n}$ with orbital index $a$ ($3z^2-r^2$ or
$x^2-y^2$) and spin projection $\sigma$, ${\bf S_n}$ is a local
spin of $t_{2g}$ electrons. Below we will consider ${\bf S_n}$ as
classical vectors. $\bm{\sigma}$ are the Pauli matrices and
$Q_{2{\bf n}}$, $Q_{3{\bf n}}$ are normal modes of vibration of
MnO$_6$ octahedron. The symbol $\langle {\bf nm}\rangle$ denotes
the summation over nearest sites. The electron part
$H_{\text{el}}$ of Hamiltonian \eqref{H} includes the kinetic
energy of $e_{g}$ electrons and the Hund's rule coupling between
the spins of $e_{g}$ and $t_{2g}$ electrons. $H_{\text{AF}}$ is
the antiferromagnetic ($J'>0$) exchange interaction between local
spins. The $H_{\text{JT}}$ term  takes into account interactions
between $e_{g}$ electrons and vibrational modes for the MnO$_6$
octahedra, here $K$ is the elastic energy and $g$ is the
electron-lattice coupling constant. The on-site Coulomb repulsion
$H_{\text{el-el}}$ includes the terms corresponding to $e_{g}$
electrons at the same and different orbitals, where the bar above
$a$ or $\sigma$ means {\it not} $a$ or {\it not} $\sigma$,
respectively.

We consider the limit $J_H\to\infty$ characteristic of manganites.
In this case, the spin of $e_g$ electrons is parallel to
$\textbf{S}_{\bf{n}}$ and we can eliminate the spin indices by the
transformation of $a_{{\bf n}a\sigma}$ to operators $c_{{\bf n}a}$
with spin projection $+1/2$ onto the direction of ${\bf S_n}$
accompanied by the transformation of hopping
amplitudes~\cite{degen}: $t^{ab}_{{\bf n}{\bf m}}\to t^{ab}_{{\bf
n}{\bf m}}\cos(\nu_{{\bf nm}}/2)$, where $\cos\nu_{{\bf nm}}={\bf
S_n}{\bf S_m}/S^2$. In addition, we assume that $t^{(l)}_{{\bf
n}{\bf m}}\to0$ for $l$ electrons, which produce maximum splitting
of $e_{g}$ level with the energy gain $-g^2/2K$, whereas $b$
electrons with non-zero hopping integrals $t^{(b)}_{{\bf n}{\bf
m}}$ produce smaller distortions of MnO$_6$ octahedra. Preliminary
calculations for the case $t^{(l)}_{{\bf n}{\bf m}}\neq0$
demonstrated that the results are not significantly affected if
the 'localized' band is much narrower than itinerant one.
Therefore, in this paper, we consider the limiting case of zero
hopping integral for 'localized' electrons. 
Hamiltonian then reads

\begin{eqnarray}\label{Heff}
H'_{\text{eff}}&=&H_{\text{eff}}-\mu\sum_{{\bf n}}\left(n_{l{\bf
n}}+n_{b{\bf
n}}\right)\,,\\
H_{\text{eff}}&=&-t\sum_{\langle{\bf nm}\rangle}c^{\dag}_{{\bf
n}}c_{{\bf
m}}\sqrt{\frac{S^2+\mathbf{S}_{\mathbf{n}}\mathbf{S}_{\mathbf{m}}}{2S^2}}-\epsilon_{\text{JT}}\sum_{{\bf
n}}n_{l{\bf n}}\nonumber\\&&+U\sum_{{\bf n}}n_{l{\bf n}}n_{b{\bf
n}}+J'\sum_{{\langle{\bf
nm}\rangle}}\mathbf{S}_{\mathbf{n}}\mathbf{S}_{\mathbf{m}}\,,\nonumber
\end{eqnarray}
where $n_{b{\bf n}}=c^{\dag}_{{\bf n}}c_{{\bf n}}$ and $n_{l{\bf
n}}=l^{\dag}_{{\bf n}}l_{{\bf n}}$ are the numbers of $b$ and $l$
electrons at site ${\bf n}$, $c^{\dag}_{{\bf n}}$, $c_{{\bf n}}$
and $l^{\dag}_{{\bf n}}$, $l_{{\bf n}}$ are the creation and
annihilation operators for the $b$ and $l$ electrons,
respectively, and $\mu$ is the chemical potential. The first three
terms in $H_{\text{eff}}$ correspond, respectively, to the kinetic
energy of $b$ electrons, JT energy of localized electrons, and
on-site Coulomb repulsion between $b$ and $l$ electrons. The last
term in $H_{\text{eff}}$ is the Heisenberg antiferromagnetic
exchange between local spins. The effective on-site Coulomb
repulsion $U$ in Eq.~\eqref{Heff} can differ from $U_2$, but has
the same order of magnitude ($\sim 5\,eV$).
$\epsilon_{\text{JT}}\sim g^2/2K$ is the JT energy gain for $l$
electrons counting from the center of $b$ electron band. The
number of localized, $n_l$, and band, $n_b$, electrons per lattice
site obeys an obvious relation $n_b+n_l=1-x$, where $x$ is the
doping level.

In our paper, we limit ourselves by the case of the large on-site
Coulomb repulsion $U$, which strongly suppresses double occupancy
of the site. Moreover, at large $U$, the characteristic time of
existence for a 'double electron state' is of the order of $1/U
\ll 1/\Omega$, where $\Omega$ is the characteristic JT phonon
frequency ($\hbar=1$), which is of the order of Debye frequency.
Therefore, the adiabatic approximation for the Jahn-Teller
distortions is applicable. As a result, the JT term in
Eq.~\eqref{Heff} is determined only by localized electrons.

Hamiltonian~\eqref{Heff} was analyzed at zero temperature in
Ref.~\onlinecite{prlKRS}. The homogenous ferromagnetic and
antiferromagnetic states as well as the phase-separated FM-AF
state were studied. The effective parameters $t$ and
$\epsilon_{\text{JT}}$ were considered to be independent of the
densities of band $n_b$ and localized $n_l$ electrons. In the
present analysis, we also neglect the dependence of parameters
$\epsilon_{\text{JT}}$ and $t$ on $n_b$ and $n_l$, but take into
account the temperature dependence of hopping integral $t$, which
can be rather strong due to polaron band
narrowing~\cite{polaron,KH}. Following Refs.
\onlinecite{polaron,KH} we write the expression for the hopping
integral in the form
\begin{equation}\label{tT}
t(T)=t_0\exp\left(-\frac{2\lambda^2}{\text{e}^{\Omega/T}-1}\right)\,.
\end{equation}
In this expression, $\lambda$ is the dimensionless electron-phonon
coupling constant. From Eq.~\eqref{tT}, it is clear that the
hopping integral $t(T)$ decreases with temperature. Even if we
take into account the finite bandwidth  for 'localized' electrons,
it will also decrease with $T$. However, the parameters describing
this behavior could be different for different bands. We cannot
assert that at high temperatures the ratio of widths for the
narrow and wide bands would be larger or smaller than at low
temperatures. In this paper, we consider the temperature range,
which is relatively small as compared to characteristic value of
$\Omega$. Thus, $t(T)$ does not vary by orders of magnitude and
the bandwidth ratio remains small.

To study the effects of temperature, we use the mean field (MF)
approximation. For this purpose, we make a decoupling procedure in
the first term of $H_{\text{eff}}$, Eq.~\eqref{Heff}, in the
following way. The values of $c^{\dag}_{{\bf n}}c_{{\bf m}}$ and
$\vartheta_{\bf
nm}\equiv\sqrt{(S^2+\mathbf{S}_{\mathbf{n}}\mathbf{S}_{\mathbf{m}})/2S^2}=
\sqrt{(1+\mathbf{e}_{\mathbf{n}}\mathbf{e}_{\mathbf{m}})/2}$ can
be written as
\begin{equation*}
c^{\dag}_{{\bf n}}c_{{\bf m}}=\left\langle c^{\dag}_{{\bf
n}}c_{{\bf m}}\right\rangle+\delta\left(c^{\dag}_{{\bf n}}c_{{\bf
m}}\right),\,\,\vartheta_{\bf nm}=\left\langle\vartheta_{\bf
nm}\right\rangle+\delta\vartheta_{\bf nm}\,,
\end{equation*}
where angular brackets mean thermal averaging,
$\mathbf{S}_{\mathbf{n}}=S\mathbf{e}_{\mathbf{n}}$, and
$\mathbf{e}_{\mathbf{n}}$ is the unit vector. Omitting the
products proportional to $\delta\left(c^{\dag}_{{\bf n}}c_{{\bf
m}}\right)\delta\vartheta_{\bf nm}$, we write the first term in
$H_{\text{eff}}$ as
$$
-\bar t\sum_{\langle{\bf nm}\rangle}c^{\dag}_{{\bf n}}c_{{\bf
m}}-\sum_{\langle{\bf nm}\rangle}\left\langle c^{\dag}_{{\bf
n}}c_{{\bf
m}}\right\rangle\left(t\sqrt{\frac{1+\mathbf{e}_{\mathbf{n}}\mathbf{e}_{\mathbf{m}}}{2}}
-\bar t\right)\,,
$$
where $\bar
t=t(T)\left\langle\sqrt{(1+\mathbf{e}_{\mathbf{n}}\mathbf{e}_{\mathbf{m}})/2}\right\rangle$.
Note that in homogenous state, $\left\langle\vartheta_{\bf
nm}\right\rangle$ does not depend on indices ${\bf n}$ and ${\bf
m}$ (for sites $\mathbf{m}$ nearest to the site $\mathbf{n}$).
Now, the effective Hamiltonian can be represented as a sum of
electronic and magnetic parts
\begin{eqnarray}
H^{MF}_{\text{eff}}=H_{\text{el}}+H_{\text{m}}-\mu\sum_{\bf{n}}(n_{l\bf{n}}+n_{b\bf{n}})\,,\nonumber\\
H_{\text{el}}=-\bar t\sum_{\langle{\bf nm}\rangle}c^{\dag}_{{\bf
n}}c_{{\bf m}}-\epsilon_{\text{JT}}\sum_{{\bf n}}n_{l{\bf
n}}+U\sum_{{\bf n}}n_{l{\bf n}}n_{b{\bf n}},\\
H_{\text{m}}=-\sum_{\langle{\bf nm}\rangle}\left[\left\langle
c^{\dag}_{{\bf n}}c_{{\bf m}}\right\rangle\left(t\vartheta_{\bf
nm}-\bar t\right)-J'S^2\mathbf{e_n}\mathbf{e_m}\right]\label{Hm}.
\end{eqnarray}

\section{Homogeneous states}

\subsection{Ferromagnetic state}

In the ferromagnetic state, we have
$\langle\mathbf{e_ne_m}\rangle=1$ far below the Curie temperature.
First, we consider the electronic sector of the problem. The
Hamiltonian $H_{\text{el}}$ is similar to that considered in
Ref.~\onlinecite{prlKRS} and the temperature $T$ enters only the
effective hopping integral $\bar{t}(T)$. To calculate the free
energy of the electronic subsystem, we use the Hubbard I
decoupling for the one-$b$-electron Green function $G_{b}({\bf
n,n}_0;\,\tau-\tau_0)=-i\langle\langle \hat{T}c_{{\bf
n}}(\tau)c^{\dag}_{{\bf n}_0}(\tau_0)\rangle\rangle$, as in
Ref.~\onlinecite{prlKRS}. Here $\tau$ is the time variable and
$\hat{T}$ is the time-ordering operator. The equation of motion
for $G_{b}({\bf n,n}_0;\,\tau-\tau_0)$ can be written in the form:
\begin{eqnarray}\label{Gb}
\left(i\frac{\partial}{\partial \tau}+\mu\right)G_{b}({\bf
n,n}_0;\,\tau-\tau_0)=\delta_{{\bf
nn}_0}\delta(\tau-\tau_0)\nonumber\\
-\bar{t}\sum_{{\bf\Delta}}G_{b}({\bf
n}+{\bf\Delta,n}_0;\,\tau-\tau_0)+U{\cal{G}}({\bf
n,n}_0;\,\tau-\tau_0)\,,
\end{eqnarray}
where summation is performed over sites nearest to site ${\bf n}$,
and ${\cal{G}}$ is the 'two-particle' Green function
${\cal{G}}({\bf n,n}_0;\,\tau-\tau_0)=-i\langle\langle
\hat{T}c_{{\bf n}}(\tau)n_{l{\bf n}}(\tau)c^{\dag}_{{\bf
n}_0}(\tau_0)\rangle\rangle$. The equation of motion for
${\cal{G}}$ is
\begin{eqnarray}\label{G4}
&&\left(i\frac{\partial}{\partial \tau}+\mu-U\right){\cal{G}}({\bf
n,n}_0;\,\tau-\tau_0)=n_{l}\delta_{{\bf
nn}_0}\delta(\tau-\tau_0)\nonumber\\
&&+i\bar{t}\sum_{{\bf\Delta}}\left\{\langle\langle\hat{T}n_{l{\bf
n}}(\tau)c_{{\bf n}+{\bf\Delta}}(\tau)c^{\dag}_{{\bf
n}_0}(\tau_0)\rangle\rangle\right.\nonumber\\
&&+\langle\langle\hat{T}l^{\dag}_{{\bf n}}(\tau)l_{{\bf
n}+{\bf\Delta}}(\tau)c_{{\bf n}}(\tau)c^{\dag}_{{\bf
n}_0}(\tau_0)\rangle\rangle\nonumber\\
&&\left.-\langle\langle\hat{T}l^{\dag}_{{\bf
n}-{\bf\Delta}}(\tau)l_{{\bf n}}(\tau)c_{{\bf
n}}(\tau)c^{\dag}_{{\bf n}_0}(\tau_0)\rangle\rangle\right\}\,.
\end{eqnarray}
The decoupling in the first term in curly brackets gives
$\langle\langle n_{l{\bf
n}}(\tau)\rangle\rangle\langle\langle\hat{T}c_{{\bf
n}+{\bf\Delta}}(\tau)c^{\dag}_{{\bf
n}_0}(\tau_0)\rangle\rangle=in_{l}G_{b}({\bf n}+{\bf\Delta},{\bf
n}_0;\,\tau-\tau_0)$. Making similar decoupling in the next two
terms, we get
\begin{eqnarray*}
&iG_{b}({\bf n,n}_0;\,\tau-\tau_0)&\\
&\times\sum\limits_{\bf\Delta}\left\{\langle\langle l^{\dag}_{{\bf
n}}(\tau)l_{{\bf
n}+{\bf\Delta}}(\tau)\rangle\rangle\right.&\left.-\langle\langle
l^{\dag}_{{\bf n}-{\bf\Delta}}(\tau)l_{{\bf
n}}(\tau)\rangle\rangle\right\}.
\end{eqnarray*}
In the homogeneous ferromagnetic state, the sum evidently
vanishes. As a result, we obtain the closed system of equations
for $G_b$ and ${\cal{G}}$. In the frequency-momentum
representation, the solutions for $G_{b}$ and ${\cal{G}}$ are as
follows
\begin{equation}\label{Gappr}
\left\{\begin{array}{rcl} G_{b}({\bf k},\omega)&=&
\displaystyle\frac{\omega+\mu-U(1-n_l)}{E_2({\bf k})-E_1({\bf
k})}\\
&\times&\left(\displaystyle\frac{1}{\omega+\mu-E_2({\bf
k})}-\frac{1}{\omega+\mu-E_1({\bf k})}\right)\,,\\&&
\\{\cal{G}}({\bf k},\omega)&=&n_l\displaystyle\frac{1+\bar
w\zeta({\bf k})G_{b}({\bf k},\omega)}{\omega+\mu-U}\,,
\end{array}\right.
\end{equation}
where
\begin{equation}\label{E12}
E_{1,2}({\bf k})=\frac{U+\bar{w}\zeta({\bf
k})}{2}\mp\sqrt{\left(\frac{U-\bar{w}\zeta({\bf
k})}{2}\right)^2+U\bar{w}\zeta({\bf k})n_l}\,.
\end{equation}
In these expressions,
\begin{equation}\label{wzt}
\bar
w=zt_0\exp\left(-\frac{2\lambda^2}{\text{e}^{\hbar\Omega/T}-1}\right)%
\left\langle\sqrt{\frac{1+\mathbf{e_0e_{\Delta}}}{2}}\right\rangle\,,
\end{equation}
and
$$
\zeta({\bf
k})=-\frac1z\sum_{\mathbf{\Delta}}\text{e}^{i\mathbf{k\Delta}}\,,
$$
where $z$ is the number of nearest neighbors. In the case of the
simple cubic lattice, we have $z=6$ and in the tight-binding
approximation
\begin{equation}\label{cubic}
\zeta({\bf k})=-\frac{1}{3}\left[\cos (k^1d)
+\cos(k^2d)+\cos(k^3d)\right],
\end{equation}
where $d$ is the lattice constant and $k^i$ are the components of
the wave vector.

From Eq.~\eqref{Gappr}, it follows that the energy spectrum of $b$
electrons includes two sub-bands given by Eq.~\eqref{E12} and the
number of states in each sub-band depends on $n_l$. In the limit
of large $U$, which is relevant to magnetic oxides,
Eq.~\eqref{E12} can be written as
\begin{equation}\label{bands}
E_1= \bar{w}\zeta({\bf k})(1-n_l),\,\,\,E_2= U+\bar{w}\zeta({\bf
k})n_l.
\end{equation}
It is clear from Eq.~\eqref{bands} that the width of the lower
sub-band is $W=2(1-n_l)\bar w$ while the width of the upper
sub-band is $2n_l\bar w$. The total number of states in two
sub-bands per site is equal to one and the number of states in the
lower and upper sub-bands is equal to $1-n_l$ and $n_l$,
respectively. Note that this result is valid for any value of $U$
as it can be demonstrated by integration of the corresponding
terms in the Green function Eq.~\eqref{Gappr}. Thus, at any doping
level $x$ and temperature $T\ll U$ the upper sub-band is empty
since $n_b+n_l=n=1-x$. In this case, it is reasonable to use the
$U\to\infty$ limit. The Green function $G_{b}$ then becomes
\begin{equation}
G_{b}({\bf k},\omega)=\frac{1-n_l}{\omega+\mu-(1-n_l)\bar
w\zeta(\mathbf{k})}\,.
\end{equation}

At low temperatures, $T\ll\bar w$, it is reasonable to represenr
the Fermi-Dirac distribution function for $b$ electrons
$f(E)=[\exp\{(E-\mu)/T\}+1]^{-1}$ by the step function
$\theta(\mu-E)$. It can be shown that this approximation works
well at $t,\epsilon_{\text{JT}}\gg J$, and $T\lesssim J$, where
$J=zJ'S^2$. An appreciable discrepancy could arise only at doping
levels $x\ll1$ and $1-x\ll1$. However, at these doping levels, the
homogeneous ferromagnetic state is unfavorable (see below).
Therefore, the number of $b$ electrons can be expressed through
the Green function as
\begin{equation}\label{nb1}
n_b=-i\int\frac{d\omega}{2\pi}\int\frac{d^3k}{(2\pi)^3}\,G_b({\bf
k},\omega+i0\!\cdot\!\text{sgn}\,\omega)\text{e}^{i\omega0}\,.
\end{equation}
This relationship defines $n_b$ as a function of chemical
potential $\mu$ and $n_l$. The value of $n_l$ depends on relative
positions of $\mu$ and $\epsilon_{JT}$.

If $\mu<-\epsilon_{\text{JT}}$, then the Jahn-Teller-induced
localization is unfavorable, $n_l=0$, $n_b=1-x$, and the chemical
potential is found from Eq.~\eqref{nb1}. With the increase of
$n_b$, $\mu$ becomes equal to $-\epsilon_{JT}$, the further growth
in the number of itinerant charge carriers is ceased, $n_l$
becomes nonzero, and $\mu$ is pinned at the level
$-\epsilon_{JT}$. In the latter case, the number of localized
electrons can be found from Eq.~\eqref{nb1} at
$\mu=-\epsilon_{JT}$ using the relation $n_b=1-x-n_l$. As a
result, we get
\begin{equation}\label{nb0}
(1-n_l)\left[1-n_0\left(-\frac{\epsilon_{JT}}{\bar
w(1-n_l)}\right)\right]=x\,,
\end{equation}
where
\begin{equation}\label{n0mu}
n_0(\mu')=\int\limits_{-1}^{\mu'}dE'\,\rho_0(E')\,,
\end{equation}
and
\begin{equation}\label{rho0}
\rho_0(E')=\displaystyle\int\frac{d^3\mathbf{k}}{(2\pi)^3}\,\delta(E'-\zeta(\mathbf{k}))
\end{equation}
is the density of states of free electrons.

At zero doping, $x=0$, the number of localized electrons $n_l=1$,
and the bandwidth $W=0$. At low doping, all electrons are
localized, $n_l=1-x$ and $W=2x\bar{w}$, until the bottom of the
band reaches the energy of $l$ electrons $-\epsilon_{\text{JT}}$
at some critical concentration $x=x_1(T)=\epsilon_{\text{JT}}/\bar
w$. At $x>x_1(T)$ the localized and band electrons coexist as long
as $x$ is smaller than the second critical doping level $x_2(T)$
at which the existence of localized electrons becomes unfavorable,
that is, $n_l=0$ if $x>x_2(T)$. From Eq.~\eqref{nb0} we find
$x_2(T)=1-n_0(-\epsilon_{\text{JT}}/\bar w)$. Naturally, such a
picture can exist only for a certain relationship between the
parameters of the model, in particular, if
$\epsilon_{\text{JT}}<\bar w$. Since $\bar w$ decreases with the
growth of temperature, both these critical concentrations, $x_1$
and $x_2$, become larger when temperature increases. Note that the
homogeneous ferromagnetic state can exist only at $x>x_1$ if
$n_b\neq 0$.

\begin{figure}
\begin{center}
\epsfig{file=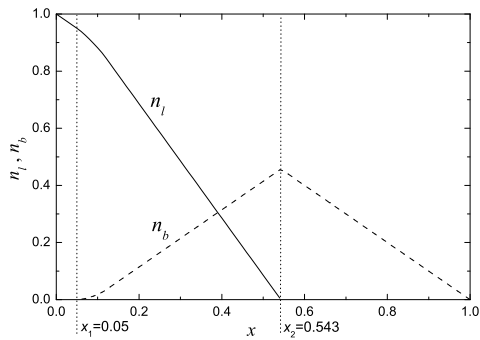,width=0.35\textwidth}
\end{center}
\caption{\label{Fignl} The dependence of $n_l$ (solid line) and
$n_b$ (dashed line) on doping concentration $x$ at
$\epsilon_{JT}/\bar w=0.05$.}
\end{figure}

The values of $n_l$ and $n_b$ as the functions of doping $x$ are
shown in Fig.~\ref{Fignl} at typical parameters of the model.
These calculations and calculations below are performed for a
simple cubic lattice with $\zeta({\bf k})$ given by
Eq.~\eqref{cubic}. In this case, the expressions for $\rho_0(E')$
and $n_0(\mu')$ in Eqs.~\eqref{rho0} and \eqref{n0mu} are found in
Appendix~\ref{SE}, see Eqs.~\eqref{rho0A} and \eqref{n0A},
respectively.

At $x_1(T)<x<x_2(T)$, the kinetic energy of $b$ electrons per site
can be written as
\begin{eqnarray}
E_{\text{kin}}&=&-\bar t\sum_{\Delta}\langle
c^{\dag}_{\bf 0}c_{{\bf\Delta}}\rangle\\
&=&-i\bar w\!\!\int\frac{d\omega}{2\pi}\int\!\!\frac{d^3k}{(2\pi)^3}\,%
\zeta(\mathbf{k})G_b({\bf
k},\omega+i0\!\cdot\!\text{sgn}\,\omega)\text{e}^{i\omega0}\nonumber
\end{eqnarray}
For further purposes we introduce the function
\begin{equation}\label{A}
A(\epsilon')=\frac1z\sum_{\Delta}\langle c^{\dag}_{\bf
0}c_{{\bf\Delta}}\rangle=-(1-n_l)\varepsilon_0\left(-\frac{\epsilon'}{1-n_l}\right)\,,
\end{equation}
where
\begin{equation}
\varepsilon_0(\mu')=\int\limits_{-1}^{\mu'}dE'\,E'\rho_0(E')\,.
\end{equation}
The kinetic energy of $b$ electrons then reads
\begin{equation}\label{Ekin}
E_{\text{kin}}=-\bar wA\left(\frac{\epsilon_{JT}}{\bar
w}\right)\,.
\end{equation}
The energy of the on-site Coulomb repulsion can be found using the
'two-particle' Green function ${\cal{G}}$
\begin{equation}
U\langle n_{b0}n_{l0}\rangle=-iU\int\frac{d\omega}{2\pi}\int\frac{d^3k}{(2\pi)^3}\,%
{\cal{G}}({\bf
k},\omega+i0\!\cdot\!\text{sgn}\,\omega)\text{e}^{i\omega0}\,.
\end{equation}
Since $\mu<U$, the pole coming from the denominator of the
function ${\cal{G}}$ (see second Eq.~\eqref{Gappr}) does not
contribute to the integral over $\omega$, and in the limit
$U\to\infty$, we get
\begin{equation}\label{Upot}
U\langle n_{b0}n_{l0}\rangle=n_l\,\bar
wA\left(\frac{\epsilon_{JT}}{\bar w}\right)\,.
\end{equation}
Note that we replace the Fermi-Dirac distribution by the step
function. Therefore, we can omit the electron entropy term and
write the free energy of electrons per site as a sum of
Eqs.~\eqref{Ekin}, \eqref{Upot}, and the JT term
\begin{equation}\label{Fel}
F_{\text{el}}=-(1-n_l)\bar wA\left(\frac{\epsilon_{JT}}{\bar
w}\right)-\epsilon_{JT}n_l\,.
\end{equation}
For $x>x_2(T)$ when $n_l=0$, we should replace $\epsilon_{JT}/\bar
w$ by $-\mu'$ in formulas \eqref{Ekin}, \eqref{Upot}, and
\eqref{Fel}, where $\mu'$ is found from the equation
$1-x=n_0(\mu')$.

Now, we consider the magnetic part of the Hamiltonian,
$H_{\text{m}}$, Eq.~\eqref{Hm}. Following the conventional mean
field approach for spin systems~\cite{Smart}, we replace
$\mathbf{e_m}=\{\sin\theta_{\mathbf{m}}\cos\phi_{\mathbf{m}},\,
\sin\theta_{\mathbf{m}}\sin\phi_{\mathbf{m}},\,\cos\theta_{\mathbf{m}}\}$
in Eq.~\eqref{Hm} by its mean value $\{0,\,0,\,m\}$, where
$m=\langle\cos\theta_{\mathbf{m}}\rangle$. As a result,
$H_{\text{m}}$ decouples into the sum of $N$ independent one-site
Hamiltonians, $H_{\text{m}}=NH_{m0}$, where
\begin{eqnarray}\label{HmMF}
H_{m0}(\cos\theta)&=&-A\left(\frac{\epsilon_{JT}}{\bar
w}\right)\left(w(T)\sqrt{\frac{1+m\cos\theta}{2}}-\bar w\right)\nonumber\\
&&+Jm\cos\theta\,,
\end{eqnarray}
$w(T)=zt(T)$. The value of $m$ is determined by the
self-consistency condition
\begin{equation}\label{Sigma}
m=\displaystyle\frac{\int\limits_{-1}^{1}du\,u\text{e}^{-\beta
H_{m0}(u)}}{\int\limits_{-1}^{1}du\,\text{e}^{-\beta
H_{m0}(u)}}\,,
\end{equation}
where $\beta=1/T$. We should also take into account that $\bar w$
is related to $H_{m0}(u)$ and $m$ by Eq.\eqref{wzt}
\begin{equation}\label{barW}
\bar w=\frac{w(T)}{\sqrt{2}}\displaystyle\frac{\int\limits_{-1}^{1}du\sqrt{1+um}\,%
\text{e}^{-\beta
H_{m0}(u)}}{\int\limits_{-1}^{1}du\,\text{e}^{-\beta
H_{m0}(u)}}\,.
\end{equation}

In order to find the Curie temperature $T_c$, we expand right hand
side of Eq.~\eqref{Sigma} in a power series in $m$. Using
Eqs.~\eqref{HmMF} and \eqref{barW} we find
$$
m=a_1(T)\beta m-a_3(T)\beta m^3-a_5(T)\beta m^5+\dots\,.
$$
The Curie temperature is found then from the condition
\begin{equation}\label{Tc}
a_1(T_C)=\frac{1}{3}\left[\frac{w(T_C)}{2\sqrt{2}}A\left(\frac{\epsilon_{JT}\sqrt{2}}{w(T_C)}\right)%
-J\right]=T_C\,.
\end{equation}
If we neglect the effect of polaron band narrowing, $\lambda=0$,
we find the explicit expression for the Curie temperature
$T_C=[w_0A(\epsilon_{JT}\sqrt{2}/w_0)/(2\sqrt{2})-J]/3$. It is
clear that polaron band narrowing reduces $T_C$.

Let us now analyze the order of the phase transition at $T=T_C$,
which depends on the sign of $a_3(T_C)$, see
Ref.~\onlinecite{Landau}. If $a_3(T_C)>0$, then $m$ tends to zero
at $T\to T_C$ as
\begin{equation}\label{Sigma2}
m(T)\approx\sqrt{\frac{a_1(T)-T}{a_3(T)}}\,.
\end{equation}
In this case, we have second order magnetic phase transition. In
the opposite case $a_3(T_C)<0$, $m$ behaves approximately as
($a_5(T)>0$)
\begin{equation}\label{Sigma1}
m(T)\approx\left.\sqrt{\frac{|a_3(T)|}{a_5(T)}+\frac{a_1(T)-T}{a_3(T)}}%
\right|_{T\to
T_C}\!\!\!\!\!\!\!\!\!\!\!\!\longrightarrow\sqrt{\frac{|a_3(T_C)|}{a_5(T_C)}}\neq0,
\end{equation}
and the transition to paramagnetic (PM) state is of the first
order. The analysis shows that $a_5(T)>0$ for any values of
parameters of the model, but the sign of $a_3(T_C)$ can be both
positive or negative depending on the doping level $x$. At some
$x=x_{12}$, the coefficient $a_3(T_C)$ changes its sign. At low
doping, $x<x_{12}$, we have $a_3(T_C)<0$ and the transition from
the FM to PM state is of the first order. At $x>x_{12}$, we have
the second order phase transition.

The free energy of the system per site is equal to
\begin{equation}\label{Ffm}
F_{\text{fm}}=F_{\text{el}}-T\ln Z_{m},\,\,%
Z_{m}=\left(S+\frac12\right)\!\int\limits_{-1}^{1}du\,\text{e}^{-\beta
H_{m0}(u)}.
\end{equation}
At $T\to0$, $m\to1$, $\bar w\to w_0=zt_0$,
$Z_{m}\approx\exp(-\beta J)$, and the free energy
$$
\left.F_{\text{fm}}\right|_{T\to0}\rightarrow-(1-n_l)w_0A\left(\frac{\epsilon_{JT}}{w_0}\right)%
-\epsilon_{JT}n_l+J\,.
$$
In the PM state, we have $m=0$, $\bar w=w(T)/\sqrt{2}$,
$Z_{m}=2S+1$, and
$$
F_{\text{pm}}=-(1-n_l)\frac{w(T)}{\sqrt{2}}A\left(\frac{\epsilon_{JT}\sqrt{2}}{w(T)}\right)%
-\epsilon_{JT}n_l-T\ln(2S+1)\,.
$$

The transition from FM to PM state does not mean that $n_b$
becomes zero. As it was mentioned above, $n_b\neq 0$ at $x>x_1(T)$
and the value of $x_1$ increases with the temperature. At a
certain temperature $T^*$, $x_1(T)$ exceeds $x$. It is clear that
$T^*>T_C$ since the FM state can exist only at $n_b\neq 0$.
Therefore, in our model, in addition to magnetic phase
transitions, there should exist an electronic transition to the
state without itinerant electrons. The temperature $T^{*}$ is
determined by the evident condition
$x_1(T^{*})=\epsilon_{JT}\sqrt{2}/w(T^{*})=x$.

\subsection{Antiferromagnetic and canted states}

In our model, there can exist other homogeneous states competing
with FM and PM states. It is natural to consider the
two-sublattice antiferromagnetic and canted states. In the canted
state, the angle $\nu$ (canting angle) between the local spins
belonging to two sublattices varies from $\pi$ (AF state) to $0$
(FM state). Here we consider AF and canting states of $G$ type,
that is, in the cubic lattice each site of one sublattice is
surrounded by sites of the second sublattice.

At finite temperatures, the canting angle is defined as
$\cos\nu=\langle\mathbf{e_0e_{\Delta}}\rangle$. In the mean field
approximation used above, we have
$\langle\mathbf{e_0e_{\Delta}}\rangle=\langle\mathbf{e_0}\rangle\langle\mathbf{e_{\Delta}}\rangle$.
Thus, we can define variable $m$ similar to that introduced in the
previous subsection, $m^2=\cos\nu$ if $\nu<\pi/2$ or
$m^2=-\cos\nu$ if $\nu>\pi/2$. In the first case, the value of
$m(T)$ and the free energy of the system is found in the same way
as for FM state from Eqs.~\eqref{HmMF}--\eqref{barW}. The only
difference is that $m=1$ at $T=0$ in FM state (or $\nu=0$) while
$m(0)<1$ (or $\nu\neq0$) in the canted state. Note that the
temperature of the transition from canted to paramagnetic state,
$T_{cant}$, is given by Eq.~\eqref{Tc}, where $T_C$ should be
replaced by $T_{cant}$.

In the case $\cos\nu<0$, instead of Eq.~\eqref{Sigma} and
\eqref{barW}, we have
\begin{equation}\label{SigmaAF}
-m=\displaystyle\frac{\int\limits_{-1}^{1}du\,u\text{e}^{-\beta
H_{m0}(u)}}{\int\limits_{-1}^{1}du\,\text{e}^{-\beta
H_{m0}(u)}}\,,
\end{equation}
\begin{equation}\label{barWAF}
\bar w=\frac{w(T)}{\sqrt{2}}\displaystyle\frac{\int\limits_{-1}^{1}du\sqrt{1-um}\,%
\text{e}^{-\beta
H_{m0}(u)}}{\int\limits_{-1}^{1}du\,\text{e}^{-\beta
H_{m0}(u)}}\,,
\end{equation}
where mean-field Hamiltonian $H_{m0}(u)$ is given by
Eq.~\eqref{HmMF}. The temperature of the phase transition from
canting to PM state is found now from the equation
\begin{equation}\label{Tcant}
\frac{1}{3}\left[J-\frac{w(T_{cant})}{2\sqrt{2}}%
A\left(\frac{\epsilon_{JT}\sqrt{2}}{w(T_{cant})}\right)\right]=T_{cant}\,.
\end{equation}
The canted state can exist at $n_b\neq 0$ since it arises in our
model due to the motion of conduction electrons.

\begin{figure}
\begin{center}
\epsfig{file=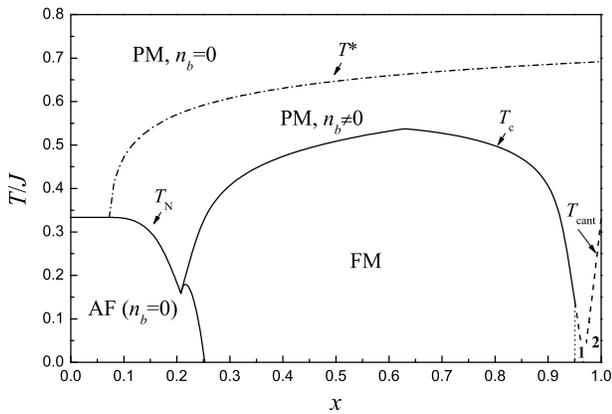,width=0.45\textwidth}
\end{center}
\caption{\label{FigHomPhDia} The phase diagram of the system
without taking into account the possibility of phase separation.
The regions $1$ and $2$ correspond to the canted states with
$\cos\nu>0$ and $\cos\nu<0$, respectively. The parameters are
$\epsilon_{JT}/w_0=0.05$, $J/w_0=0.01$, $\lambda=10$, and
$\Omega/w_0=0.03$.}
\end{figure}

At doping levels $x<x_1(T)$, we have $n_b=0$ and the AF ordering
is favorable. At rather low doping, $x<x_1(0)$, $n_b=0$ at any
temperature. In this case, the N\'eel temperature is independent
of $x$ and is determined by Eq.~\eqref{Tcant} as $T^{0}_{N}=J/3$.
At higher doping, there can occur a transition from the AF state
to the PM state with $n_b\neq 0$, and the N\'eel temperature,
$T_N$, is determined by the comparison of free energies of
corresponding states. In this doping range, the AF state turns out
to be more favorable than the canted state. With the further
increase of $x$, the phase with $n_b\neq 0$ has the lower energy,
but this phase is FM rather than canted. The temperature of the
AF--FM transition is found from the comparison of corresponding
free energies. This transition is of the first order. Note that
the phase diagram exhibits AF--FM--PM triple point at $x=x_3$
($x_3<x_{12}$). The canted state can exist in the doping range
$1-x\ll1$, where the number of itinerant electrons is too small to
stabilize the FM state. The corresponding phase diagram in the
$x-T$ plane is shown in Fig.~\ref{FigHomPhDia}. We see that the
competition between the itinerant and localized electrons gives
rise to a rather complicated magnetic phase diagram of the system.

\section{Phase separation}

Up to this point, we considered only homogeneous states, however,
it is well known that different inhomogeneous states are possible
in the systems with strongly correlated electrons. So, we should
compare the free energies of the states studied in the previous
sections with those for inhomogeneous states. As a typical example
of an inhomogeneous state, we analyze here the droplet model of
electronic phase separation widely discussed in connection with
manganites and other magnetic oxides. Among the possible types of
phase separation, we treat below the coexistence of different
phases: AF--FM, FM--PM, and PM phases with different values of
$n_b$. For simplicity, we do not include into consideration the
phase separation involving the canted state since it exists only
in the narrow doping range. Note that our model always leads to
such kind of phase separation, where we have $n_b=0$ in one of the
phases.

We consider a system separated into two phases with the volume
concentrations $p$ and $1-p$. In the homogeneous phases, the
electron concentration per site coincides with the doping level
$x$. In the inhomogeneous states, the electrons can be
redistributed between the regions with different phases. Let
$n_b\neq 0$ in the first (F) phase and $n_b=0$ in the second (A)
phase, the electron density per site in the first phase is $x_f$
and in the second phase is $x_a$. So, doping level $x$ lies
between $x_a$ and $x_f$. The charge conservation requires
$p\,x_f+(1-p)x_a=x$.

\begin{figure}
\begin{center}
\epsfig{file=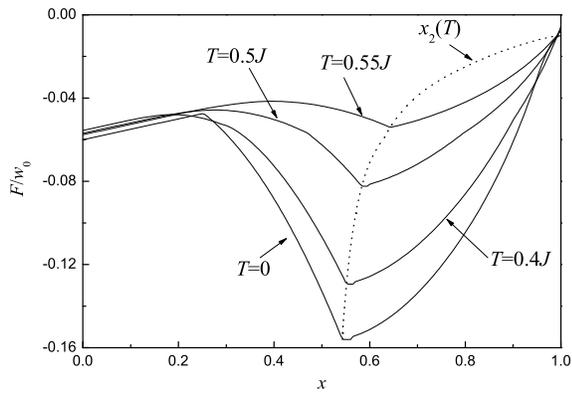,width=0.42\textwidth}
\end{center}
\caption{\label{FigFEnergy} Free energy of homogeneous state
$F_{\text{hom}}=\min(F_{\text{fm}},\,F_{\text{af}},\,F_{\text{pm}},\,F_{\text{cant}})$
{\it vs.} doping level $x$, at different temperatures. The dot
curve corresponds to the function $x_2(T)$. The parameters are
$\epsilon_{JT}/w_0=0.05$, $J/w_0=0.01$, $\lambda=10$, and
$\Omega/w_0=0.03$.}
\end{figure}

In Fig.~\ref{FigFEnergy}, we show the dependence of the free
energy of the most favorable homogeneous state
$F_{\text{hom}}=\min\left(F_{\text{fm}},F_{\text{af}},F_{\text{pm}},F_{\text{cant}}\right)$
on the doping level at different temperatures. The
$F_{\text{hom}}(x)$ curves have two minima: one at $x=0$ and
another near $x=x_2(T)$. Then we could expect that $x_a$ should be
around zero, while $x_f$ should be close to $x=x_2(T)$.

The phase separation corresponds to the non-uniform charge density
and we should take into account the Coulomb contribution to the
total energy. This contribution depends on the structure of
inhomogeneous state. To evaluate the Coulomb energy, we assume the
spherical geometry of the phase-separated state. Namely, at
$p<0.5$, the sample is modelled as an aggregate of spheres of F
phase embedded into A matrix or that of A spheres in F phase for
$p>0.5$. For this geometry, it is reasonable to calculate the
Coulomb energy using the Wigner-Seitz approximation: each F or A
sphere of radius $R_s$ is surrounded by a spherical cell of radius
$R_{cell}$, such as the volume of the cell is $4\pi
R_{cell}^3/3=V/N_s$, where $V$ is the volume of the sample and
$N_s$ is the number of spheres. The radius $R_{cell}$ is related
to $R_s$ as $R_s=p^{1/3}R_{cell}$ for $p<0.5$, and
$R_s=(1-p)^{1/3}R_{cell}$ for $p>0.5$. The total electric charge
inside this cell is zero and the Coulomb energy of the system is
the sum of the electrostatic energies of these cells. Following
Ref.~\onlinecite{Lor}, we obtain the expression for the Coulomb
energy per site $E_c$ at $p<0.5$
\begin{equation}\label{Ec}
E_c=\frac{2\pi
e^2}{5\epsilon d}\,\left(x_f-x_a\right)^2\left(\frac{R_s}{d}\right)^2%
p\left(2-3p^{1/3}+p\right)\,,
\end{equation}
where $\epsilon$ is the average permittivity of the sample and $d$
is the lattice constant. In the case $p>0.5$, we should replace
$x_f\leftrightarrow x_a$ and $p\to1-p$.

The $b$ electrons in the phase-separated state are confined within
a restricted volume. The corresponding size quantization gives
rise to the change in the density of states. The additional
contribution to the energy (per site) is proportional to the total
surface area between F and A phases, and at $p<0.5$ can be written
in the form
\begin{equation}\label{Es}
E_{s}=p\,\frac{3d}{R_s}\sigma(x_f)\,,
\end{equation}
where surface energy $\sigma(x_f)$ is calculated in
Appendix~\ref{SE}. In the case $p>0.5$, we should change
$p\to1-p$.

The Coulomb \eqref{Ec} and surface \eqref{Es} contributions to the
total energy depend on the size $R_s$ of the inhomogeneities.
Minimization of $E_{cs}=E_c+E_s$ with respect to $R_s$ gives at
$p<0.5$
\begin{equation}\label{Rs}
R_s=d\left(\frac{15\sigma(x_f)}{4\pi%
u(x_f-x_a)^2\left(2-3p^{1/3}+p\right)}\right)^{1/3}\!\!\!\!\!\!,
\end{equation}
\begin{equation}
E_{cs}=3\left(u\frac{9\pi}{10}(x_f-x_a)^2\sigma(x_f)^2\right)^{1/3}\!\!\!\!\!\!%
p\left(2-3p^{1/3}+p\right)^{1/3}\!\!\!\!\!\!.
\end{equation}
where $u=e^2/\epsilon d$.

Let us estimate the parameter $u$ and the characteristic size of
inhomogeneities $R_s$. Using typical values of parameters for
manganites $z=6$, $t_0=0.3$\,eV, $d=0.4$\,nm, and $\epsilon=20$ we
find $u\approx0.18$\,eV, and $u/w_0\approx0.15$. The surface
energy $\sigma(x)$ is calculated in Appendix~\ref{SE}. For
example, at $x=0.2$ the optimization procedure gives $x_a\ll
x_f\approx 0.5$, $p\approx 0.4$ and $\sigma(x_f)\approx 1.4\cdot
10^{-2}w_0$ and from Eq.~\eqref{Rs} we find $R_s\approx1.5\,d$,
that is, the inhomogeneity contains
$N_{s}=4\pi(R_s/d)^3/3\approx15$ unit cells. Thus, we see that
$N_s\gg 1$ for characteristic values of parameters and the
Wigner-Seitz approximation is applicable. However, this
approximation overestimates the Coulomb contribution because of a
sharp boundary of the inhomogeneities. Therefore, the above values
for $R_s$ and $N_s$ could be considered as lower estimates.

In the phase-separated state, to find the values of $x_f$ and
$x_a$ at given $x$ and $T$, it is necessary to minimize the total
free energy
\begin{equation}\label{Fps}
F_{\text{PS}}(x_f,x_a)=p\,F_{\text{F}}(x_f)+(1-p)F_{\text{A}}(x_a)+E_{cs}(x_f,x_a,p)
\end{equation}
with respect to $x_f$ and $x_a$, where $p=(x-x_a)/(x_f-x_a)$.

At temperature $T>T_{max}^*=T^*(x=1)$ (see Fig.~\ref{FigHomPhDia})
only homogeneous PM state with $n_b=0$ is possible. As it follows
from the numerical and analytical analyzes of the free energy,
below $T_{max}^*$, the function of two variables
$F_{\text{PS}}(x_f,x_a)$ has two minima if $x<x_2(T)$ for any
phases F and A within the considered hierarchy of parameters
($J\ll \epsilon_{JT}<w_0$). The first minimum at the point
$x_a=x_f=x$ corresponds to some homogeneous state while the second
minimum at the point $x_a=0$, $x_f=x_2(T)$ corresponds to the
phase-separated state. In the limit $u\to0$, the second minimum is
the global minimum of the function $F_{\text{PS}}(x_f,x_a)$ at
$0<x<x_2(T)$ and $T<T^{*}_{\text{max}}$, and the PS state is
favorable. When $u$ increases, the range of phase separation in
the plane $(x,T)$ gradually narrows and disappears at some
critical value $u_c$. Note that there is no localized electrons in
a more metallic F phase ($x_f=x_2(T)$), and vice versa, in the
insulating A phase $n_b(T)=0$ since $x_a=0$, as it was mentioned
in connection to Fig.~\ref{FigFEnergy}.

Since the concentrations $x_a$ and $x_f$ are independent of the
doping level, the temperatures of magnetic phase transitions in
both phases do not depend on $x$. The N\'eel temperature of the AF
phase is $T^0_N=J/3$, whereas the Curie temperature for the FM
phase can be found from the equation
\begin{equation}\label{TcPS}
\frac{1}{3}\left[-\frac{w(T_C)}{2\sqrt{2}}\,\varepsilon_0%
\left(-\frac{\epsilon_{JT}\sqrt{2}}{w(T_C)}\right)-J\right]=T_C\,.
\end{equation}
The N\'eel and Curie temperatures in the PS state calculated in
such a way correspond to the macroscopic phases with the size of
inhomogeneities $R_s/d\gg1$. If $R_s\sim d$, the values of $T_N$
and $T_C$ in the PS state can differ from those calculated above.

\begin{figure}
\begin{center}
\epsfig{file=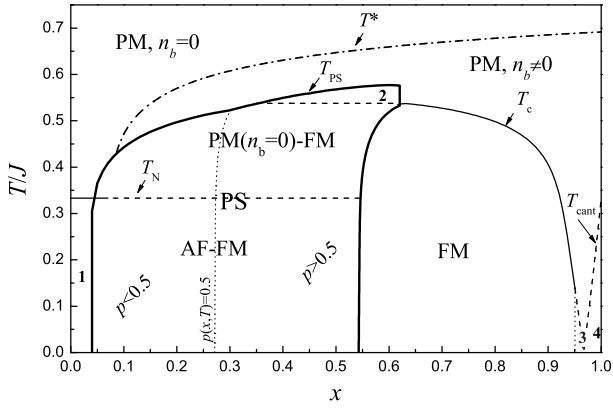,width=0.45\textwidth}
\end{center}
\caption{\label{FigPSPhDia} The phase diagram of the model at
$\epsilon_{JT}/w_0=0.05$, $J/w_0=0.01$, $\lambda=10$,
$\Omega/w_0=0.03$, and $u/w_0=0.5$. The numbers denote: 1.
homogeneous AF phase; 2. the mixture of two PM states with
$n_b\neq 0$ and $n_b=0$; 3. and 4. homogeneous canted states.}
\end{figure}

The region where the PS state is favorable can be found from the
comparison of free energy $F_{\text{PS}}(x_2(T),0)$ with the free
energies of homogeneous states at given $x$ and $T$. The phase
diagram of the system in the $(x,\,T)$ plane is shown in
Fig.~\ref{FigPSPhDia}. The range of existence for the PS state is
bounded by the curve $T_{\text{PS}}(x)$. In the PS state, the
content $p$ of the metallic ($n_b\neq 0$) phase varies with the
temperature and the doping level. Hence, the insulator-metal
transition is possible when $p$ exceeds the percolation threshold.

\begin{figure}
\begin{center}
\epsfig{file=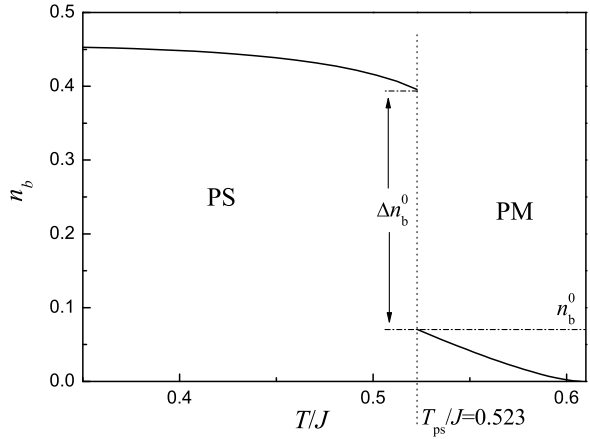,width=0.43\textwidth}
\end{center}
\caption{\label{FigNbT} The temperature dependence of $n_b$ at
$x=0.3$, $\epsilon_{JT}/w_0=0.05$, $J/w_0=0.01$, $\lambda=10$,
$\Omega/w_0=0.03$, and $u/w_0=0.5$.}
\end{figure}

Let us now discuss the transition of the system from the PS to a
homogeneous state. The volume fraction of the F phase in the PS
state is $p(T)=x/x_2(T)$. Depending on the relation between the
temperatures $T_{\text{PS}}$, $T_C$, $T_N$, and $T^{*}$, the
system can pass from the PS state to the FM ($p=1$), AF ($p=0$),
and PM (with $n_b\neq0$ or $n_b=0$) homogeneous states. In all
cases, the number of itinerant electrons $n_b$ undergoes a sudden
change at the transition to the homogeneous state. The temperature
dependence of $n_b$ is shown in Fig.~\ref{FigNbT}.

\section{Effect of magnetic field}

In this section, we consider the effect of magnetic field on the
properties of the system. We take into account only the effect of
the magnetic field on the local spin. This corresponds to the
limit of classical local spin  $S\gg 1$. Thus, in the presence of
external DC magnetic field $\mathbf{H}$, we should add the term
$-\mu_{B}g\sum_{\mathbf{n}}\mathbf{S_n}\mathbf{H}$ in
Hamiltonian~\eqref{Heff}, where $\mu_{B}$ is the Bohr magneton and
$g$ is the Lande factor. As a result, the magnetic field term
modifies only the magnetic Hamiltonian, Eq.~\eqref{Hm},
\begin{eqnarray}\label{HmH}
H_{\text{m}}=-\sum_{\langle{\bf nm}\rangle}\left[\left\langle
c^{\dag}_{{\bf n}}c_{{\bf
m}}\right\rangle\left(t\sqrt{\frac{1+\mathbf{e}_{\mathbf{n}}\mathbf{e}_{\mathbf{m}}}{2}}
-\bar t\right)\right.\\
\left.\phantom{\frac{H_{\text{m}}}{2}}-J'S^2\mathbf{e_n}\mathbf{e_m}+\mathbf{e_nh}\right],%
\,\,\,\,\,\,\,\mathbf{h}=\mu_{B}gS\mathbf{H}\,.\nonumber
\end{eqnarray}

In the FM state, the one-site magnetic Hamiltonian~\eqref{HmMF}
corresponding to the MF approximation takes the form
\begin{eqnarray}\label{HmMFH}
H_{m0}(\cos\theta)&=&-A\left(\frac{\epsilon_{JT}}{\bar
w}\right)\left(w(T)\sqrt{\frac{1+m\cos\theta}{2}}-\bar w\right)\nonumber\\
&&+Jm\cos\theta-h\cos\theta\,,
\end{eqnarray}
where the direction of magnetic field is parallel to $z$ axis. The
mean value $m=\langle S^{z}_{0}\rangle/S$ is found by solving the
system of equations~\eqref{Sigma} and \eqref{barW} with
Hamiltonian \eqref{HmMFH}. At $T<T_C$, the correction to the free
energy in the presence of magnetic field $\delta F\sim-h$, whereas
in paramagnetic phase $\delta F\sim-h^2/T_C$.

In the AF or canted states, the result depends on the mutual
orientation of $\mathbf{h}$ and the vector
$\mathbf{l}=(\langle\mathbf{e}_0\rangle-\langle\mathbf{e_{\Delta}}\rangle)/2$.
The minimum of the free energy corresponds to the case
$\mathbf{h}\bot\mathbf{l}$. Let the vector $\mathbf{l}$ be
parallel to $z$ axis and the magnetic field $\mathbf{h}$ be
parallel to the $x$ axis. The mean values of the directions of
local spins in two sublattices, $\langle\mathbf{e}_{0}\rangle$ and
$\langle\mathbf{e_{\Delta}}\rangle$ can be written as
$\langle\mathbf{e}_{0}\rangle=\{m,\,0,\,l\}$,
$\langle\mathbf{e_{\Delta}}\rangle=\{m,\,0,\,-l\}$, where $m$ is
proportional to the magnetization of the system. The one-site
magnetic Hamiltonian then has the form
\begin{eqnarray}\label{HmMFHAF}
H_{m0}(\theta,\phi)&=&-A\left(\frac{\epsilon_{JT}}{\bar
w}\right)\left(w(T)\sqrt{\frac{1+\mathbf{e}_0\langle\mathbf{e_{\Delta}}\rangle}{2}}-\bar
w\right)\nonumber\\
&&+J\mathbf{e}_0\langle\mathbf{e_{\Delta}}\rangle-h\sin\theta\cos\phi\,,
\end{eqnarray}
where
$\mathbf{e}_0=\{\sin\theta\cos\phi,\,\sin\theta\sin\phi,\,\cos\theta\}$
and
$$
\mathbf{e}_0\langle\mathbf{e_{\Delta}}\rangle=-l\cos\theta+m\sin\theta\cos\phi\,.
$$
The values of $l$ and $m$ are found from the equations
\begin{equation}\label{lm}
\left\{\begin{array}{rcl} l&=&\displaystyle\frac{\int
d\mathbf{\Omega}\,\cos\theta\,\text{e}^{-\beta
H_{m0}(\theta,\phi)}}{\int d\mathbf{\Omega}\,\text{e}^{-\beta
H_{m0}(\theta,\phi)}}\\ \\
m&=&\displaystyle\frac{\int
d\mathbf{\Omega}\,\sin\theta\cos\phi\,\text{e}^{-\beta
H_{m0}(\theta,\phi)}}{\int d\mathbf{\Omega}\,\text{e}^{-\beta
H_{m0}(\theta,\phi)}}\,.
\end{array}\right.
\end{equation}
Expression for the effective bandwidth $\bar{w}$~\eqref{barWAF}
now takes the form
\begin{equation}\label{barWH}
\bar w(H,T)=\frac{w(T)}{\sqrt{2}}\displaystyle\frac{\int
d\mathbf{\Omega}\sqrt{1+\mathbf{e}_0\langle\mathbf{e_{\Delta}}\rangle}\,\,%
\text{e}^{-\beta H_{m0}(\theta,\phi)}}{\int
d\mathbf{\Omega}\,\text{e}^{-\beta H_{m0}(\theta,\phi)}}\,.
\end{equation}

At high magnetic fields $h\sim T_N\sim J$, we get from
Eq.~\eqref{lm} that $l=0$, and the system passes from the AF or
canted state to the FM one. The typical N\'eel temperature in
manganites is $T_N\sim100$\,K, and the value of $h=J$ corresponds
to the fields of the order of $H\sim100$\,T.

If $h/J\ll1$, the system~\eqref{lm} can be solved perturbatively.
The correction to the free energy of the AF state is $\delta
F\sim-h^2/T_N$ both below and above $T_N$, because there is no
spontaneous magnetization in the system at $H=0$. The external
magnetic field favors the FM state in comparison to the AF and
canted states. In particular, it reduces the temperature of
transition from the AF to FM state. The magnetic field leads to
the increase in the effective hopping integral $\bar t$ due to the
alignment of local spins and thus to the growth in the number of
$b$ electrons and the value of $T^{*}$. At the transition from the
PS to PM homogeneous state the magnetic field results in the
increase of the transition temperature, and the difference $\Delta
T_{\text{PS}}=T_{\text{PS}}(H)-T_{\text{PS}}(0)\sim h$. At the
transition from the PS to FM state, $\Delta T_{\text{PS}}$ can be
both positive and negative depending on the parameters.

\begin{figure}
\begin{center}
\epsfig{file=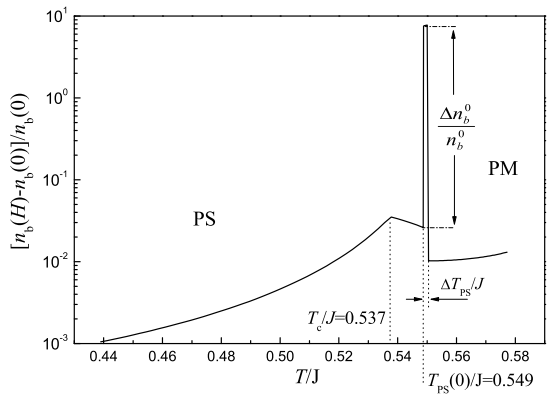,width=0.4\textwidth}
\end{center}
\caption{\label{FigDNbT} The temperature dependence of
$(n_b(h)-n_b(0))/n_b(0)$ at $h/J=0.2$, $x=0.3$,
$\epsilon_{JT}/w_0=0.05$, $J/w_0=0.01$, $\lambda=10$,
$\Omega/w_0=0.03$, and $u/w_0=0.3$.}
\end{figure}

As was mentioned above, the number of itinerant electrons $n_b$
differs significantly below and above $T_{\text{PS}}$. Therefore,
the shift of the transition temperature $T_{\text{PS}}$ with the
magnetic field gives rise to a significant change in the number of
itinerant electrons. The temperature dependence of the ratio
$(n_b(H)-n_b(0))/n_b(0)$ near the transition temperature
$T_{\text{PS}}(0)$ is shown in Fig.~\ref{FigDNbT}. The narrow peak
in this ratio is a manifestation of the step in $n_b(T)$ shown in
Fig.~\ref{FigNbT}. Since the position $T_{\text{PS}}$ of the step
in $n_b(T)$ depends on the magnetic field, a small change in $H$
causes a significant change in the number of charge carriers at a
fixed temperature near $T_{\text{PS}}$. The number of itinerant
charge carriers determines the value of metallic conductivity of
the system. Thus, the large change of $n_b$ in magnetic field can
be related to the colossal magnetoresistance effect.

\section{Conclusions}

We discussed a ''minimal model''  dealing with the competition
between the localization and metallicity in manganites. The
Hamiltonian of the model takes into account the essential physics
of strongly correlated electron systems with the Jahn-Teller ions:
it is, in fact, the Hubbard model with the strong electron-lattice
interaction, the Hund's rule intraatomic coupling, and AF
interatomic exchange between local spins. Such an approach
provides a possibility to understand the difference between the
number of itinerant charge carriers $n_b$ and the doping
level~\cite{prlKRS}. It is shown that $n_b$ can be significantly
lower than the number of the charge carriers implied by the doping
level. The models of similar type were discussed in
Refs.~\onlinecite{Rama,Rama1,Pai}. However, the possibility of the
phase separation was not considered in these papers.

Here we demonstrate that in the framework of our model phase
separation can exist in a wide range of intermediate doping
concentrations disappearing at low and high doping level. These
predictions are in agreement with the general features of the
experimentally found phase diagrams of
manganites~\cite{dagbook,Dag,Nag}. The obtained results suggest
the existence of the droplet-type of the electronic phase
separation that was widely discussed in literature (see, e.g.,
Ref.~\onlinecite{dagbook}). We calculated the relative content of
different phases in the phase separated states and found the size
of such droplets (ferrons). For the characteristic values of
parameters, a droplet includes 10--30 unit cells. These results
could, in particular, serve as a key to an adequate description of
the transport properties of manganites that could not be done in
the framework of the single-band models~\cite{prbMi,zhetf04}.

As it was mentioned above, various models related to the strongly
correlated electrons exhibit a pronounced tendency toward phase
separation. The effective Hamiltonian of our model~\eqref{Heff}
is, in fact, a generalization of the Falicov-Kimball
model~\cite{fal}. The latter model describes a system with the
hybridization of an electronic band and a localized level. The
Falicov-Kimball model is often used as a toy model in the analysis
of heavy-fermion materials and it also leads to the phase
separation phenomena~\cite{PSfal}. So, we believe that our
approach could be applicable not only to manganites but also to a
wider class of strongly correlated electron systems. Note that the
analogy between the Falicov-Kimball model and the Hamiltonian of
the $s-d$ type with the Jahn-Teller interaction was indicated in
Ref.~\onlinecite{Pai} but in the case without phase separation.

In this paper, we analyzed the phase diagram of the model in the
$x-T$ plane. The effect of temperature manifests itself mainly in
the change of effective hopping integral $t$ due to the polaron
band narrowing and the entropy term in the free energy due to
thermal fluctuations of local spins. The polaron band narrowing is
described by standard formula \eqref{tT}, see
Refs.~\onlinecite{polaron,KH}. The behavior of the local spins was
treated using the mean field approximation. We find that at low
temperatures the system is in a state with a long-range magnetic
order: AF, FM or AF--FM phase separated state. We demonstrate that
at high temperatures there can exist two types of the paramagnetic
state, a usual one with $n_b=0$ and that with $n_b\neq 0$. In the
intermediate temperature range, the phase diagram includes
different kinds of the PS states: AF--FM, FM--PM, and PM with
different content of itinerant electrons.

The applied magnetic field leads to the changes in the phase
diagram. It evidently favors the FM ordering and, consequently, to
the increase of the number of itinerant electrons. The effect of
the magnetic field was analyzed accounting for the alignment of
the local spins in the applied magnetic field.

It is demonstrated that in our model the metal-insulator
transition can take place at some characteristic values of the
doping $x$ corresponding to the crossover between different kinds
of the phase separation. It can be induced by changing temperature
or the magnetic field and is of a percolation type. This
transition can be related to the colossal magnetoresistance
effect.

Note that in the present treatment we assume that the effective
parameters $t$ and $\epsilon_{\text{JT}}$ do not depend on the
doping level $x$. To verify the applicability of such
approximation, we calculated the phase diagram with $t$ and
$\epsilon_{\text{JT}}$ depending linearly on $x$. We found that
the phase diagram remains qualitatively the same even if $t$ and
$\epsilon_{\text{JT}}$ vary by the factor 2--3 provided the
hierarchy of the model parameters ($J\ll \epsilon_{JT}<w_0$)
remains unchanged.

Note also that we included to our analysis the long-range Coulomb
interaction related to the macroscopic charge redistribution in
the phase separated state. It allowed us to estimate the size of
the inhomogeneities. At the same time, we did not take explicitly
into account the corresponding terms in the model
Hamiltonian~\eqref{Heff}. However, if we would like to consider
the effect of charge ordering, we have to include at least the
nearest-neighbor Coulomb repulsion.

Here we considered only the ''minimal model'' describing the
effect of the phase separation. Therefore, we did not include into
consideration the possibility of the charge ordering to focus the
discussion on the interplay between localized and itinerant
electrons. It is not impossible to include the charge ordering to
the model of such type. The first attempt was made in
Ref.~\onlinecite{Rama1}, but the possibility of phase separation
was not considered there. The effect of the charge ordering could
change the results for $x$ near 0.5. Therefore, it is reasonable
to consider only the range $x<0.5$, if we want to compare our
predictions with the actual situation in the doped magnetic
oxides.

\section*{Acknowledgments}

The work was supported by the Russian Foundation for Basic
Research, project No. 05-02-17600 and by the Russian Presidential
Grant No. NSh-1694.2003.2.


\appendix

\section{Surface energy}\label{SE}

In this section, we calculate the surface energy coming from the
size quantization. The expression for the free energy of $b$ and
$l$ electrons can be written in terms of the density of states in
the form (see Eqs.~\eqref{A} and \eqref{Fel})
\begin{equation}\label{FelS}
F_{\text{el}}=\bar
w(1-n_l)^2\int\limits_{-1}^{\mu'}dE'\,E'\rho_0(E')-\epsilon_{JT}n_l\,,
\end{equation}
where
$$
\mu'=-\frac{\epsilon_{JT}}{\bar w(1-n_l)}\,.
$$
This expression is valid for $x<x_2(T)$ and $\mu=-\epsilon_{JT}$.
The density of states for the system of itinerant electrons in the
volume $V$ is
\begin{equation}\label{rho}
\rho_0(E')=\frac{1}{V}\sum_{{\bf n}}\,\delta\left(E'-\zeta({\bf
k_n})\right)\,,
\end{equation}
where momentum ${\bf k}$ varies over a discrete set of values,
depending on boundary conditions and geometry of the system. The
function $\zeta({\bf k_n})$ is normalized to unity, that is
$|\zeta({\bf k_n})|\leq1$. In the thermodynamic limit
$V\to\infty$, the sum in Eq.~\eqref{rho} can be replaced by the
integral over ${\bf k}$ in the first Brillouin zone multiplied by
$V/(2\pi)^3$. At finite $V$, we derive an approximate expression
for the density of states in the case of cubic lattice,
corresponding a small value of $\Delta=Sd/V$, where $S$ is the
surface area and $d$ is the lattice constant.

Let the sample has the shape of a parallelepiped $P$ with sides
$L_1$, $L_2$, and $L_3$ (in units of the lattice constant $d$).
The Dirichlet boundary conditions for the conduction electron wave
function $\psi_{\bf m}$ is used, that is, $\psi_{\bf m}=0$ for
${\bf m}\,\nexists\,P$. In this case, the momentum ${\bf k_n}$
takes values ${\bf k_n}=\pi {\bf n}/(L_{\alpha}+1)$, where
$n^{\alpha}=1,\,2\dots\,L_{\alpha}$ ($\alpha=1,\,2,\,3$). For
large $L_{\alpha}$, we can use the trapezium rule for the sum over
$k^{\alpha}$
$$
\sum_{n=1}^{L_{\alpha}}f(k^{\alpha}_{n})=\frac{L_{\alpha}+1}{\pi}\!\!
\int\limits_{0}^{\pi}\!\!dk\,f(k)-\frac12(f(0)+f(\pi))+O\left(\frac{1}{L_{\alpha}}\right).
$$
As a result, for 3D sum over ${\bf n}$ we obtain
\begin{eqnarray}\label{Sum1}
&\sum\limits_{\bf n}&f({\bf
k_n})\simeq\left(V+\frac{S}{2}\right)\!\!
\int\!\!\frac{d^3k}{\pi^3}\,f({\bf
k})\\
&-&\frac{V}{2}\left(
\int\!\!\frac{d^2k}{\pi^2L_1}\left[f(\{0,k^2,k^3\})+f(\{\pi,k^2,k^3\})\right]\right.\nonumber\\%
&+&\!\!\!\!\int\!\!\frac{d^2k}{\pi^2L_2}\left[f(\{k^1,0,k^3\})+f(\{k^1,\pi,k^3\})\right]\nonumber\\
&+&\left.\!\!\int\!\!\frac{d^2k}{\pi^2L_3}\left[f(\{k^1,k^2,0\})+f(\{k^1,k^2,\pi\})\right]\right)\,,\nonumber
\end{eqnarray}
where 3D and 2D momentum integrations are performed
in the range $0\leq k^{\alpha}\leq\pi$, and
$S=2(L_1L_2+L_1L_3+L_2L_3)$ is the surface area of the
parallelepiped in the units of lattice constant.

Using formula~\eqref{Sum1} with $f({\bf k})=\delta(E'-\zeta({\bf
k}))$, we can calculate the density of states.
Relation~\eqref{Sum1} can be simplified for the case of cubic
symmetry since
$\zeta(k^1,k^2,k^3)=\zeta(k^2,k^1,k^3)=\zeta(k^3,k^2,k^1)$. In the
absence of external fields, we have $\zeta({\bf k})=\zeta(-{\bf
k})$, and the integration in Eq.~\eqref{Sum1} can be extended to
$-\pi\leq k^{\alpha}\leq\pi$. As a result, we obtain for the
density of states
\begin{widetext}
\begin{equation}\label{rhoD}
\rho(E')=\left(1+\frac{\Delta}{2}\right)\rho_0(E')
-\frac{\Delta}{4}\int\frac{d^2p}{(2\pi)^2}\,%
\left[\delta\left(E'-\zeta(\{0,p^1,p^2\})\right)+%
\delta\left(E'-\zeta(\{\pi,p^1,p^2\})\right)\right]\,,
\end{equation}
\end{widetext}
where $\rho_0(E')$ is the density of states at $V\to\infty$. Note,
that the density of states in the form~\eqref{rhoD} depends only
on the ratio $S/V$ and does not depend on the shape of the sample.
We believe, that Eq.~\eqref{rhoD} is applicable for any geometry
of the system, provided that the minimum linear dimension $L$ is
large compared to the lattice constant (see, for
example~\cite{BB,Ivan}).

Let us calculate now the surface energy for the spectrum in the
tight-binding approximation for the simple cubic lattice,
$\zeta({\bf k})=-[\cos(k^1d)+\cos(k^2d)+\cos(k^3d)]/3$. In the
limit $V\to\infty$, the formulas for the density of states
$\rho_0(E')$, density of itinerant electrons $n_0(\mu)$, and their
kinetic energy $\varepsilon_0(\mu)$ can be written in the
following form
\begin{eqnarray}\label{rho0A}
\rho_0(E')&=&\int\limits_{0}^{+\infty}\frac{du}{\pi}\,J_0^3\left(\frac{u}{3}\right)%
\cos(E'u)\,,\\
\label{n0A}n_0(\mu')&=&\int\limits_{0}^{+\infty}\frac{du}{\pi}\,J_0^3\left(\frac{u}{3}\right)
\frac{\sin(u)+\sin\left(\mu'u\right)}{u}\,,\\
\label{E0A}\varepsilon_0(\mu')&=&\int\limits_{0}^{\infty}\frac{du}{\pi}\,J_0^3\left(\frac{u}{3}\right)%
\left[\frac{\mu'\sin(\mu'u)-\sin(u)}{u}\right.\nonumber\\
&&\left.+\frac{\cos(\mu'u)-\cos(u)}{u^2}\right]\,,
\end{eqnarray}
where $J_0$ is the Bessel function. Now, Eq.~\eqref{rhoD} can be
rewritten as
\begin{eqnarray}\label{rhoD2}
\rho(E')&=&\left(1+\frac{\Delta}{2}\right)\rho_0(E')\\
&&-\frac{\Delta}{4}\left(\rho^{(2)}_0(E'+1/3)+\rho^{(2)}_0(E'-1/3)\right)\,,\nonumber
\end{eqnarray}
where
\begin{equation}
\rho^{(2)}_0(E')=\int\limits_{0}^{+\infty}\frac{du}{\pi}\,J_0^2\left(\frac{u}{3}\right)%
\cos(E'u)
\end{equation}
is the density of states in the 2D case. The number $n_l$ and the
free energy of electrons (in the case $n_l\neq0$) is given by
Eqs.~\eqref{nb0} and \eqref{FelS}, where instead of $\rho_0(E')$
we should use the density of states Eq.~\eqref{rhoD2}. Note that
$n_l$ and $F_{\text{el}}$ are functions of $\Delta$. In the
considered limit $\Delta\ll 1$, we can use the perturbation
technique to calculate the surface energy $\sigma$. Representing
the number of $l$ electrons and the free energy $F_{\text{el}}$ in
the form $n_l=n_l^{(0)}+\Delta\,n_l^{(1)}+\dots$,
$F_{\text{el}}=F_{\text{el}}^{(0)}+\Delta\,\sigma+\dots$, and
expanding Eqs.~\eqref{nb0}, \eqref{n0mu}, and \eqref{FelS}, one
obtains
\begin{eqnarray}
n_l^{(1)}&=&-\frac12\left(1-n_l^{(0)}\right)\\%
&&\times\frac{n_0(\mu_0')-\frac12\left(n_0^{(2)}(\mu_0'+\frac{1}{3})+n_0^{(2)}(\mu_0'-\frac{1}{3})\right)}%
{1-n_0(\mu_0')+\mu_0'\rho_0(\mu_0')}\,,\nonumber
\end{eqnarray}
\begin{eqnarray}
\sigma&=&\bar w\left\{(1-n_l^{(0)})\left(\mu_0'^2\rho_0(\mu_0')-2\varepsilon_0(\mu_0')\right)-\frac{\epsilon_{\text{JT}}}{\bar w}\right\}n_l^{(1)}\nonumber\\
&+&\left\{2\varepsilon_0(\mu_0')-\varepsilon^{(2)}_0(\mu_0'+\frac13)-\varepsilon^{(2)}_0(\mu_0'-\frac13)\right.\\
&+&\left.\frac{1}{3}\left[n_0^{(2)}(\mu_0'+\frac13)-n_0^{(2)}(\mu_0'-\frac13)\right]\right\}%
\frac{\bar w(1-n_l^{(0)})^2}{4},\nonumber
\end{eqnarray}
where $n_l^{(0)}$ is determined by Eq.~\eqref{nb0} at $\Delta=0$,
$\mu_0'=-\epsilon_{\text{JT}}/[\bar w(1-n_l^{(0)})]$,
$n_0^{(2)}(\mu')$ and $\varepsilon^{(2)}_0(\mu')$ are the number
of $b$ electrons and their energy in 2D case at $\Delta=0$,
\begin{eqnarray}\label{nE20}
n_0^{(2)}(\mu')&=&\int\limits_{0}^{+\infty}\frac{du}{\pi}\,J_0^2\left(\frac{u}{3}\right)%
\frac{\sin\left(\frac{2u}{3}\right)+\sin(\mu'u)}{u}\,,\\
\varepsilon_0^{(2)}(\mu')&=&\int\limits_{0}^{\infty}\frac{du}{\pi}\,J_0^2\left(\frac{u}{3}\right)%
\left[\frac{\mu'\sin(\mu'u)-\frac{2}{3}\sin\left(\frac{2u}{3}\right)}{u}\right.\nonumber\\
&&\left.+\frac{\cos(\mu'u)-\cos\left(\frac{2u}{3}\right)}{u^2}\right]\,.
\end{eqnarray}

\begin{figure}
\begin{center}
\epsfig{file=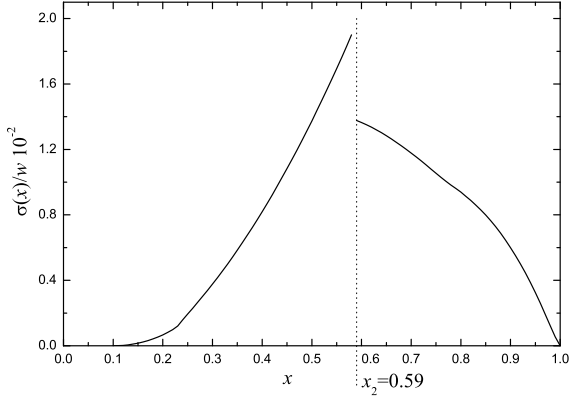,width=0.42\textwidth}
\end{center}
\caption{\label{FigSigma} The surface energy {\it vs} doping
concentration $x$ at $\epsilon_{\text{JT}}/\bar w=0.1$.}
\end{figure}

At doping concentration $x>x_2(T)$ when $n_l=0$, we should use the
equation $1-x=n_0(\mu')$ for the chemical potential, where
$\rho_0(E')$ is substituted by $\rho(E')$. As a result, we obtain
the expression for the surface energy
\begin{eqnarray}
\sigma&=&\frac{\bar w\mu_0'}{4}\left\{n_0^{(2)}(\mu_0'+\frac{1}{3})%
+n_0^{(2)}(\mu_0'-\frac{1}{3})-2n_0(\mu_0')\right\}\nonumber\\
&+&\frac{\bar w}{2}\left\{\varepsilon_0(\mu_0')-\frac12\left[%
\varepsilon^{(2)}_0(\mu_0'+\frac{1}{3})+\varepsilon^{(2)}_0(\mu_0'-\frac{1}{3})\right]\right.\nonumber\\%
&+&\left.\frac{1}{6}\left[n_0^{(2)}(\mu_0'+\frac{1}{3})-n_0^{(2)}(\mu_0'-\frac{1}{3})\right]\right\},
\end{eqnarray}
where $\mu_0'$ is found from the equation $1-x=n_0(\mu_0')$ at
$\Delta=0$. The dependence $\sigma(x)$ is shown in
Fig.~\ref{FigSigma}. The function $\sigma(x)$ is discontinuous at
$x=x_2(T)$. This singularity stems from the kink in the free
energy $F_{\text{el}}(x)$ (see Fig.~\ref{FigFEnergy}).

In the approximation under discussion, the corresponding
corrections to the magnetic contribution to the free energy
$F_{m}$ are of the order of $\Delta^2$, and therefore we omit
them.


\begin{references}

\bibitem{DagSci}
E.~Dagotto, Science {\bf 309}, 257 (2005); New J. Phys. {\bf 7},
67 (2005).

\bibitem{NagSup}
E.\,L.~Nagaev, Usp. Fiz. Nauk {\bf 165}, 529
(1995)[Physics-Uspekhi {\bf 38}, 497 (1995)].

\bibitem{fermion}
A.\,H.~Castro Neto and B.\,A.~Jones \prb {\bf 62}, 14975 (2000);
P. Chandra, P. Coleman, J.\,A.~Mydosh, and V.~Tripathi, Nature
(London) {\bf 417}, 831 (2002).

\bibitem{dagbook}
E. Dagotto, {\em Nanoscale Phase Separation and Colossal
Magnetoresistance: The Physics of Manganites and Related
Compounds} (Springer-Verlag, Berlin, 2003).

\bibitem{cobalt}
P.\,L. Kuhns, M.\,J.\,R. Hoch, W.\,G.~Moulton, A.\,P.~Reyes,
J.~Wu, and C.~Leighton, \prl {\bf 91}, 127202 (2003).

\bibitem{nickel}
G.~Blumberg, M.\,V.~Klein, and S.-W.~Cheong, \prl {\bf 80}, 564
(1998); R.~Lemanski, J.\,K.~Freericks, and G.~Banach, \prl {\bf
89}, 196403 (2002).

\bibitem{low-d}
H. Seo, C. Hotta, and H.~Fukuyama, Chem. Rev., {\bf 104}, 5005
(2004).

\bibitem{Nag67}
E.\,L. Nagaev, Pis'ma v ZhETF {\bf 6}, 484 (1967) [JETP Lett. {\bf
6}, 18 (1967)].

\bibitem{Kasuya}
T. Kasuya, A. Yanase, and T.~Takeda, Solid State Commun. {\bf 8},
1543, 1551 (1970).

\bibitem{Bala}
A.\,M. Balagurov, V.\,Yu. Pomjakushin, D.\,V.~Sheptyakov,
V.\,L.~Aksenov, P.~Fischer, L.~Keller, O.\,Yu.~Gorbenko,
A.\,R.~Kaul, and N.\,A.~Babushkina, \prb {\bf 64}, 024420 (2001).

\bibitem{Ueh}
M. Uehara, S. Mori, C.\,H.~Chen, and S.-W. Cheong, Nature (London)
{\bf 399}, 560 (1999).

\bibitem{KKK}
M.\,Yu. Kagan, K.\,I. Kugel, and D.\,I.~Khomskii, Zh. Teor. Eksp.
Fiz. {\bf 120}, 470 (2001) [JETP {\bf 93}415 (2001)].

\bibitem{stripes}
S. Mori, C.\,H. Chen, and S.-W.~Cheong, Nature (London) {\bf 392},
473 (1998); P.\,G.~Radaelli, D.\,E.~Cox, L.~Capogna, S.-W.~Cheong,
and M.~Marezio, \prb {\bf 59}, 14440 (1999); B.~Raveau,
M.~Hervieu, A.~Maignan, and C.~Martin, J. Mater. Chem. {\bf 11},
29 (2001).

\bibitem{KuKh}
D.\,I.~Khomskii and K.\,I.~Kugel, \prb {\bf 67}, 134401 (2003).

\bibitem{Kak}
M.\,Yu. Kagan and K.\,I. Kugel, Usp. Fiz. Nauk. {\bf 171}, 577
(2001) [Physics - Uspekhi {\bf 44}, 553 (2001)].

\bibitem{t-J}
V.\,J. Emery, S.\,A. Kivelson, and H.\,Q.~Lin, \prl {\bf 64}, 475
(1990).

\bibitem{Viss}
P.\,B. Visscher, \prb {\bf 10}, 943 (1974).

\bibitem{PSfal} J.\,K.~Freericks, Ch.~Gruber, and N.~Macris, \prb {\bf 60}, 1617
(1999); J.\,K.~Freericks, E.\,H.~Lieb, and D.~Ueltschi, \prl {\bf
88}, 106401 (2002); M.\,M.~Ma\'ska and K.~Czajka, phys.~stat.~sol.
(b) {\bf 242}, 479 (2005).

\bibitem{zhao}
J.\,H. Zhao, H.\,P. Kunkel, X.\,Z.~Zhou, and G.~Williams, \prb
{\bf 66}, 184428 (2002).

\bibitem{prbMi} A.\,L.~Rakhmanov, K.\,I.~Kugel, Ya.\,M.~Blanter, and
M.\,Yu.~Kagan, \prb {\bf 63}, 174424 (2001).

\bibitem{zhetf04}
K.\,I. Kugel, A.\,L.~Rakhmanov, A.\,O. Sboychakov, M.\,Yu.~Kagan,
I.\,V.~Brodsky, and A.\,V.~Klaptsov, Zh. Eksp. Teor. Fiz. {\bf
125}, 648 (2004) [JETP {\bf 98}, 572 (2004)].

\bibitem{prlKRS}
K.\,I. Kugel, A.\,L. Rakhmanov, and A.\,O.~Sboychakov, \prl {\bf
95}, 267210 (2005).

\bibitem{Millis} A.\,J.~Millis, P.\,B.~Littlewood, and B.\,I.~Shraiman,
\prl {\bf 74}, 5144 (1995).

\bibitem{hor} J.~Bala, P.~Horsch, and F.~Mack, \prb {\bf 69},
094415 (2004).

\bibitem{bus} M.~Gulacsi, A.~Bussmann-Holder, and A.\,R.~Bishop,
\prb {\bf 71}, 214415 (2005).

\bibitem{Rama} T.\,V.~Ramakrishnan, H.\,R.~Krishnamurthy, S.\,R.~Hassan,
and G.\,V.~Pai, \prl {\bf 92}, 157203 (2004).

\bibitem{Rama1}
O.~C\'epas, H.\,R.~Krishnamurthy, and T.\,V.~Ramakrishnan, \prl
{\bf 94}, 247207 (2005).

\bibitem{Good}
J.\,B. Goodenough, {\em Magnetism and the Chemical Bond}
(Interscience, New York, 1963).

\bibitem{Zener}
C. Zener, Phys. Rev. {\bf 82}, 403 (1951).

\bibitem{Pai} G.\,V.~Pai, S.\,R.~Hassan,
H.\,R.~Krishnamurthy, and T.\,V.~Ramakrishnan, Europhys. Lett.
{\bf 64}, 696 (2003).

\bibitem{degen} P.\,G. de Gennes, Phys. Rev. {\bf 118}, 141 (1960).

\bibitem{polaron}
A.\,S.~Alexandrov and N.\,F.~Mott, \emph{Polarons and Bipolarons}
(World Scientific, Singapore, 1995).

\bibitem{KH} K.\,I.~Kugel and D.\,I.~Khomskii, Zh. Eksp. Teor. Fiz.
{\bf 79}, 987 (1980) [Sov. Phys. JETP {\bf 52}, 501 (1980)].

\bibitem{Smart}
J.\,S. Smart, {\em Effective Fields Theories of Magnetism},
(Saunders, London, 1966).

\bibitem{Landau}
L.\,D.~Landau and E.\,M.~Lifshitz, {\em Statistical Physics}
(Butterworth-Heinemann, Oxford, 1980), Part 1.

\bibitem{Lor} J.~Lorenzana, C.~Castellani, and C.~di~Castro,
Europh. Lett. {\bf 57}, 704 (2002).

\bibitem{Dag}
E. Dagotto, T. Hotta, and A.~Moreo, Phys. Reports {\bf 344}, 1
(2001).

\bibitem{Nag}
E.\,L. Nagaev, Phys. Reports {\bf 346}, 387 (2001).

\bibitem{fal} L.\,M.~Falicov and J.\,C.~Kimball, \prl {\bf 22}, 997 (1969).

\bibitem{BB} R.~Balian and C.~Bloch, Ann. Phys. (N.Y.) {\bf 60}, 401
(1970).

\bibitem{Ivan} I.~Gonz\'alez, J.~Castro, and D.~Baldomir, Phys.
Lett. A {\bf 298}, 185 (2002).

\end{references}
\end{document}